\documentclass[10pt]{article}
\pdfoutput=1
\usepackage{cite}
\usepackage{amsmath,bbm}
\usepackage{amssymb}
\usepackage{graphicx} \graphicspath{{figures/}}
\usepackage[latin1]{inputenc}
\usepackage{mathrsfs}
\usepackage{mathtools}
\usepackage[dvipsnames]{xcolor}
\usepackage[footnotesize]{caption}
\usepackage{xcolor}
\usepackage{hyperref}
\usepackage[hmarginratio=1:1,top=32mm,columnsep=20pt]{geometry}
\usepackage{paralist}
\usepackage{abstract}
\usepackage{upgreek}
\usepackage{slashed}
\usepackage{multirow}

    \makeatletter
    \let\@fnsymbol\@arabic
    \makeatother

\newcommand{\bea}{\begin{eqnarray}}
\newcommand{\eea}{\end{eqnarray}}
\newcommand{\be}{\begin{equation}}
\newcommand{\ee}{\end{equation}}
\newcommand{\ba}{\begin{array}}
\newcommand{\ea}{\end{array}}

\def\gsim{\mathrel{\rlap{\lower4pt\hbox{\hskip1pt$\sim$}}
    \raise1pt\hbox{$>$}}}

\title{\vspace{-15mm}
	\fontsize{16pt}{10pt}\selectfont
	\textbf{Prospects for Heavy Scalar Searches at the LHeC}
	}	
\author{%
	\large
	\textsc{Luigi~Delle~Rose$^\star$, %\foonote{E-Mail: \texttt{XXX}	,  
	Oliver~Fischer$^\dagger$, %\footnote{E-mail: \texttt{oliver.fischer@unibas.ch}}
	A. Hammad$^\circ$
	}\\[10pt]
	\normalsize	$^\star$ INFN, Sezione di Firenze, and Department of Physics and Astronomy, University of Florence, \\
	\normalsize	Via G. Sansone 1, 50019 Sesto Fiorentino, Italy \\[5pt]
	\normalsize	$^\dagger$ Institute for Nuclear Physics, Karlsruhe Institute of Technology, \\
	\normalsize	Hermann-von-Helmholtz-Platz 1, D-76344 Eggenstein-Leopoldshafen, Germany \\[5pt]
	\normalsize	$^{\circ}$ Department of Physics, University of Basel,\\
	\normalsize	Klingelbergstr.~82, CH-4056 Basel, Switzerland
	\vspace{-5mm}
	}
\date{}

\begin{document}
\maketitle

\begin{abstract}
In this article we study the prospects of the proposed Large Hadron electron Collider (LHeC) in the search for heavy neutral scalar particles.
We consider a minimal model with one additional complex scalar singlet that interacts with the Standard Model (SM) via mixing with the Higgs doublet, giving rise to a SM-like Higgs boson $h_1$ and a heavy scalar particle $h_2$. Both scalar particles are produced via vector boson fusion and can be tested via their decays into pairs of SM particles, analogously to the SM Higgs boson. 
Using multivariate techniques we show that the LHeC is sensitive to $h_2$ with masses between 200 and 800 GeV down to scalar mixing of $\sin^2 \alpha \sim 10^{-3}$.
\noindent 

\end{abstract}

\section{Introduction}
One of the most important tasks of high-energy particle physics of our time is the detailed measurement of the couplings of the recently discovered Higgs boson.
Thanks to the recent progress at the Large Hadron Collider (LHC) some of the Higgs boson properties are known with $\sim 5\%$ precision. 
After the high-luminosity LHC most of the so-called $\kappa$ parameters, which quantify the Higgs signal strength in terms of its Standard Model (SM) prediction, may be known with a precision between 3\% and 5\%  with mostly equal contribution from statistical and systematic uncertainty \cite{Cepeda:2019klc}.
Since the Higgs boson provides a notorious portal to new physics, these measurements leave plenty of room for physics beyond the SM in the scalar sector.

A new generation of machines is presently discussed, which are to measure the Higgs boson properties with high precision. 
Among others it is worth mentioning the so-called Higgs factories, the high-luminosity electron-positron colliders at 250 GeV \cite{Gomez-Ceballos:2013zzn,Baer:2013cma,CEPC-SPPCStudyGroup:2015csa}.

It is possible, however, to improve the precision of many Higgs boson measurements at the LHC already, by upgrading the collider with a new linear electron recovery linac \cite{Angal-Kalinin:2017iup}, and colliding one of the proton beams with the new electron beam at about 60 GeV.
The resulting facility is the Large Hadron electron Collider (LHeC) \cite{Bruening:2013bga} which can be operated concurrently to the LHC at $\sim$1.2 TeV centre-of-mass energy with a total integrated luminosity of 1 ab$^{-1}$.
One of its prime objectives is the improvement of the PDF sets which would ameliorate many LHC studies \cite{Klein:2016uwv}.
This can be expected to significantly reduce the PDF-associated systematic uncertainties of LHC Higgs precision studies and also provide a very important input for many exotic BSM studies \cite{Klein:2016uwv}. 
Also a measurement of the main decay mode of Higgs boson into the $b\bar b$ final state could be improved from $\sim 6\%$ to about 1\% level precision.
Moreover, fig.\ 3 of ref.\ \cite{Bruning:2652313} shows that the $\kappa_{Z}$ and $\kappa_{W}$ measurements can be improved from the 2 to 3 percent level at the HL-LHC to the subpercent level when combined with LHeC measurements.
The LHeC also has been shown to be more than competitive with Higgs boson measurements, cf.\ \cite{Han:1985zn,Han:2009pe,Kumar:2015kca,Tang:2015uha} and to bring unique opportunities with respect to Beyond the SM (BSM) physics, cf.\ e.g.\ \cite{Cakir:2009xi,Liang:2010gm,Zhang:2015ado,Antusch:2016ejd,Curtin:2017bxr}. 

Owing to the importance of measuring the Higgs sector, many LHC analyses are searching for additional neutral scalar bosons which can be produced and decay via their mixing with the SM Higgs boson. 
Like most BSM studies at the LHC, these searches have to deal with very high rates of SM backgrounds.\footnote{
Even BSM searches with clean signatures like, e.g.,\ multi-lepton, or same-sign dilepton final states, have to contend with large additional background sources due to pile up, mis identification, and towering QCD background rates.
}
These analyses can access the squared scalar mixing angle on the order of $10\%$ and are particularly sensitive to heavy scalars with masses above a few hundred GeV.
It is interesting to note that there are some hints in the LHC data that can be interpreted as a 270 GeV neutral heavy scalar \cite{vonBuddenbrock:2015ema,vonBuddenbrock:2016rmr,vonBuddenbrock:2017gvy} which was referred to as the ``Madala hypothesis''  \cite{vonBuddenbrock:2017jqp}.
These observations, together with the background-related limitations of the LHC, represent our main motivation to study the prospects of searches for heavy neutral scalars at the LHeC.
A similar study on a preparatory level, motivated by the ``Madala hypothesis'', has been performed in \cite{Mosomane:2017jcg}.

In this paper we present a detailed study for searching a heavy Higgs-like at the LHeC via its dominant decays into SM gauge bosons, i.e, $WW$ or $ZZ$, which in turn give rise to the final states $4l$, $2l2j$ or $\nu l 2j$, among others. Compared with existing analyses at LHC, (cf.\ \cite{Khachatryan:2015cwa}) we find plenty of room for a discovery.

\section{The Model}
Extended Higgs sectors are ubiquitous in many BSM scenarios and, as such, represent a very strong case for experimental searches. 
In particular, extra scalars that are singlets under the SM gauge group are envisaged in some of the most natural extensions of the SM, 
ranging from Supersymmetry to Composite Higgs models and to GUT extensions (in which the new scalar sector provides the mass to the extra massive gauge bosons). Moreover, in some scenarios the SM singlet scalars can act as a portal to SM neutral fields in dark sectors which otherwise would remain completely unobserved. 
Extra scalar singlets also play a major role in models of electroweak baryogenesis as they represent one of the most economical possibilities to realize a first-order electroweak phase transition. 

\begin{figure}
\centering
\includegraphics[width=0.4\textwidth]{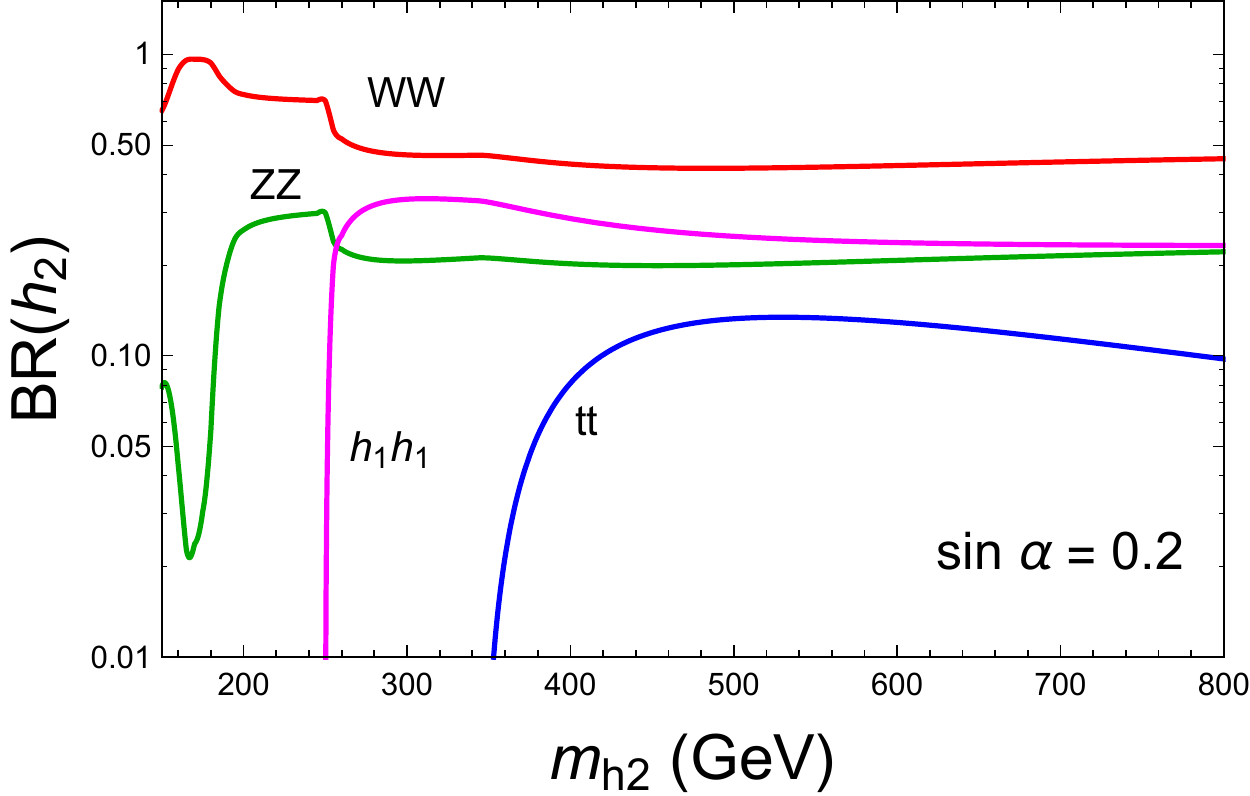}
\caption{Dominant branching ratios of the heavy Higgs boson $h_2$ as a function of its mass for a fixed value of the scalar mixing angle, $\sin \alpha=0.2$.}
\label{fig:BR}
\end{figure}

Motivated by the aforementioned scenarios, we consider a simple extension of the SM with a complex neutral scalar boson $S$, singlet under the SM gauge group.
The scalar sector is thus described by the potential
\bea
\label{eq:scalarpotential}
V(H,S) = m_1^2 H^\dag H + m_2^2 \, S^\dag S + \lambda_1 (H^\dag H)^2 + \lambda_2 (S^\dag S)^2 + \lambda_3 (H^\dag H)(S^\dag S),
\eea
which is the most general renormalizable scalar potential of the SM $SU(2)$ Higgs doublet $H$ and the complex scalar $S$. 
The mass eigenstates from the resulting mass matrix correspond to the physical fields $h_{1,2}$, which are given by
\bea
\left( \begin{array}{c} h_1 \\ h_2 \end{array} \right) = \left( \begin{array}{cc} \cos \alpha & - \sin \alpha \\  \sin \alpha & \cos \alpha \end{array} \right)  \left( \begin{array}{c} H  \\ S \end{array} \right),
\eea 
where the scalar mixing angle $\alpha$ and  the masses of the physical scalars are defined in terms of the original parameters of the potential as
\bea
\label{eq:ScalarAngle}
\tan 2 \alpha &=& \frac{\lambda_3 \, v \, x}{\lambda_1 \, v^2 - \lambda_2 \, x^2} \,, \\
\label{eq:ScalarMasses}
m_{h_{1,2}}^2 &=& \lambda_1 v^2 + \lambda_2 x^2 \mp \sqrt{\left( \lambda_1 v^2 - \lambda_2 x^2\right)^2 + \left( \lambda_3 v x \right)^2} \,,
\eea
with $m_{h_{2}} > m_{h_{1}}$ and $h_1$ identified with the 125 GeV Higgs boson. In the previous equations, $v = 246.22$ GeV and $x$ are the vacuum expectation values of the $H$ and $S$ fields, respectively. \\
After mass mixing, the mass eigenstate $h_1$ corresponds to the SM-like Higgs, by which we mean that in the limit of the scalar mixing angle $\alpha \to 0$ we recover the Higgs boson with the interactions and properties as predicted by the SM.
The scalar mixing yields couplings between the mass eigenstate $h_2$ and the SM fermions and gauge bosons proportional to those of a SM Higgs boson of the same mass, with a rescaling factor given by $\sin \alpha$. 
On top of the SM-like interactions, the last term in eq.\ \eqref{eq:scalarpotential} proportional to $\lambda_3$ gives rise to a coupling between $h_1$ and $h_2$, which yields e.g.\ the additional decay channel $h_2 \to 2 h_1$ if $m_{h_2}>2m_{h_1}$. If no other decay modes are available for the $h_2$, as it is the case for our simple setup, the phenomenology of the $h_2$ below the $h_1 h_1$ threshold is similar to that of the SM Higgs with $m_h = m_{h_2}$, with same the branching ratios and a total decay width simply rescaled by $\sin^2 \alpha$. Above the threshold, the branching ratios of $h_2$ into SM final states is given by
\bea
\textrm{BR}(h_2 \to \textrm{SM}) = \textrm{BR}_\textrm{SM}(h_2 \to \textrm{SM}) (1 - \textrm{BR}(h_2 \to h_1 h_1))
\eea
with $\textrm{BR}_\textrm{SM}(h_2 \to \textrm{SM})$ being the SM one. The branching ratio $\textrm{BR}(h_2 \to h_1 h_1)$ is computed from the corresponding partial decay width which can be expressed explicitly as
\bea
 \Gamma( h_2 \rightarrow h_1 h_1) &=& \left( \frac{ \sin 2 \alpha}{v} (\cos \alpha + \frac{v}{x} \sin \alpha)  (\frac{m_{h_2}^2}{2} + m_{h_1}^2) \right)^2 \frac{1}{32 \pi \, m_{h_2}} \left( 1 -  \frac{4 m_{h_1}^2}{m_{h_2}^2} \right)^{1/2} \,.
\eea
The main branching ratios of the heavy scalar $h_2$ are shown in fig.\ \ref{fig:BR} as a function of $m_{h_2}$ for a scalar mixing angle $\sin \alpha = 0.2$ and $x \gg v$. 
In BSM scenarios in which the SM gauge group is extended by extra abelian gauge factors as in the $B-L$ models, or unified into a larger simple group as in GUT extensions, the vev $x$ of the extra scalar participates in the spontaneous symmetry breaking patterns and sets the scale of the mass of the extra gauge bosons. As the void LHC searches of such states have pushed the corresponding masses to the TeV range or above, 
it is reasonable to assume a large separation between $x$ and the electroweak scale.

\begin{figure}
\centering
\begin{minipage}{0.49\textwidth}
\centering
\includegraphics[width=0.8\textwidth]{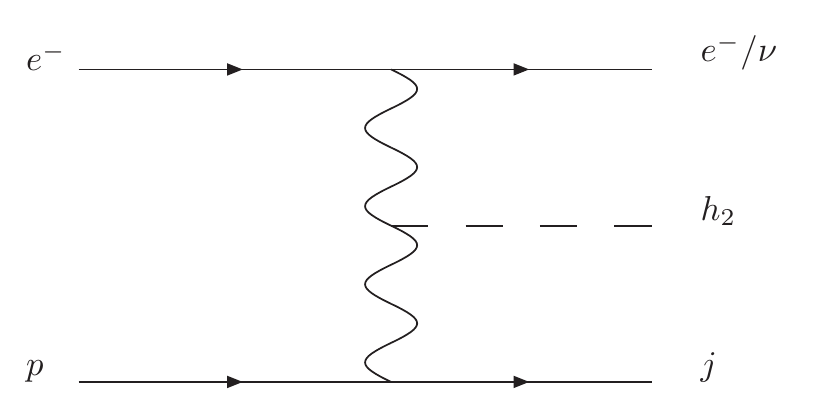}
\end{minipage}
\begin{minipage}{0.49\textwidth}
\centering
\includegraphics[width=0.8\textwidth]{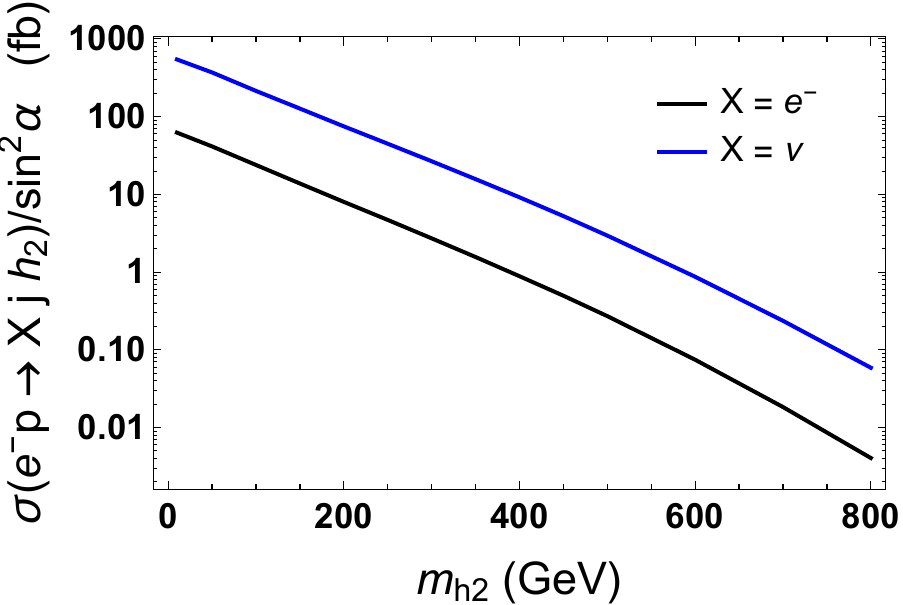}
\end{minipage}
\caption{Heavy Higgs production cross section from the process $e^- p  \to X\,j\,h_2$ at the LHeC. The final state $X$ being an electron ($e^-$) and a neutrino ($\nu$) denotes the neutral current (NC) and charged current (CC) interaction, respectively.}
\label{fig:xsections}
\end{figure}

In electron-proton collisions at the LHeC, the heavy Higgs boson $h_2$ can be produced through vector boson fusion (VBF), via the charged (CC) or neutral currents (NC), cf. the left panel of fig.\ \ref{fig:xsections}. The resulting cross sections for this process, normalised by $\sin^2 \alpha$, are shown in the right panel of fig.\ \ref{fig:xsections}.
It is worth pointing out that the CC cross section is larger by about an order of magnitude, as expected due to the small coupling of the Z boson to the charged leptons, which affects the choice of final states for the search strategies below.

\section{Heavy Higgs search strategy}
In the following we investigate the prospects of the LHeC in the search for a heavy Higgs $h_2$ in the mass region $m_{h_2}>m_{h_1}$ by focusing on its leading decay modes into SM weak gauge bosons, $WW$ and $ZZ$.
It is worth mentioning that the other interesting decay channels which would be extremely useful to characterise the phenomenology of the extra scalar, and to eventually discriminate among different models, are the di-higgs and di-top decay modes. Nevertheless, here we consider only those search channels which will most likely represent the priorities of the research program for heavy scalars at the LHeC.

\subsection{Signatures and analysis}
In particular, the following signatures will be studied:
\begin{enumerate}
\item $\mu_{\ell\ell}^Z:= h_2\to ZZ \to 4\ell$
\item $\mu_{\ell q}^Z:=h_2\to ZZ \to 2\ell2q$
\item $\mu_{\ell q}^{W}:= h_2 \to WW \to \nu\ell 2j $
\end{enumerate}
As already stated above, further channels exist, which are not expected to add significantly to the final sensitivity of the LHeC to heavy Higgses but are also very interesting in their own right. Examples for these channels are the boosted mono-$Z$ ($h_2 \to Z_{qq}Z_{\rm inv}$), di-top ($h_2\to t \bar t$) and di-Higgs ($h_2\to 2h_1$), and $\ell \ell' + E_{\rm miss}$.
The all hadronic final state $V_{qq}V_{qq} \to 4j$ adds significantly to the signal statistics, but it brings complication via the nature of the jet reconstruction and the additional combinatorics that are necessary to identify the beam remnant jet. Since we expect the complications to cancel out the advantage in statistics (at least partially) we omit studying this channel here.

Since the signal channels 1, 2, and 3 consist of two vector bosons of high invariant mass, we consider diboson processes in the SM as our primary background. 
Additional backgrounds exist in the form of processes with initial- and final state radiation of electrons and gluons. These can have large cross sections and mimic the signal final state. Due to the typically small momenta (particularly in the transverse direction) we neglect those contributions in the following.
Furthermore, we also consider signal and background processes only at the leading order and neglect corrections at higher order, which may lead to small modifications in the cross sections and also in the kinematic shapes of the observables.

The background processes considered here and the corresponding cross sections are listed in tab.\ \ref{tab:bkg}.
For these backgrounds we use a systematic uncertainty of $2\%$ \cite{Bruening:2013bga}.

The centre-of-mass at the LHeC is boosted with respect to the laboratory system due to the asymmetric beam energies which pushes the final states towards positive $\eta$ values. 
Accordingly, for heavy $h_2$, which requires larger parton energies, the decay products are strongly forward boosted with large (positive) $\eta$ values. 
This provides a good handle to separate signal from background events. 
For $h_2$ masses that become comparable to the centre-of-mass energy of $\sim1.2$ TeV this good separability is countered by the reduction of the total cross section due to the restricted phase space. \\
For the reconstruction of the signal we require that the beam-remnant jet from the deep inelastic scattering (DIS) off the proton has a transverse momentum of $P_T(j) > 10$ GeV and a pseudo rapidity of $|\eta(j)| < 4.5$ for geometric acceptance, while for leptons we require $P_T(l) > 2$ GeV and $|\eta(l)| < 4.5$, with $l$ accounts for electrons and muons.
These threshold values are representative for studies at the LHeC \cite{REF Missing!!!}.

For the simulation of the signal and background event samples, the Monte Carlo event generator MadGraph5 version 2.4.3~\cite{Alwall:2014hca} is employed. 
As usual, parton shower and hadronisation is taken care of by Pythia6~\cite{Sjostrand:2006za} while the fast detector simulation is carried out by Delphes~\cite{deFavereau:2013fsa}. We use the Delphes detector card from the LHeC collaboration.
We note that Pythia needs to be patched \cite{UtaKlein} in order to achieve a reasonable event generation efficiency and that it is crucial that the first (second) beam, as inputted in the MadGraph run card, corresponds to the proton (electron) to correctly match the asymmetric detector setup implemented in the Delphes card.
We use an electron beam of 60 GeV with 80\% polarisation, the proton beam with energy of 7 TeV and we consider a total integrated luminosity of 1/ab.
It is important to notice, that a smaller electron beam energy would result in a smaller production cross section of the $h_2$ and it would also reduce the LHeC reach with respect to the heavy Higgs mass due to the more restricted phase space.

We perform the analysis for five benchmark masses $m_{h_2}$ from 200 GeV to 800 GeV and, for illustrative purposes, we present detailed results for a specific benchmark point, allowed by current LHC searches \cite{Sirunyan:2018qlb}, defined by $m_{h_2} = 500$ GeV and $\sin \alpha = 0.2$.
 For larger heavy Higgs masses, the number of events drops significantly and, consequently, the error bands highly enlarge over the expected median preventing us from reaching a reasonable statistical conclusion. 
All the backgrounds from tab.\ \ref{tab:bkg} with all possible decay channels were included for each of the three signal channels, $4\ell,\, 2\ell + 2j, \, \ell + 2j + \slashed{E}_T$. 
A total of $10^7$ events for each signal and background sample was simulated. \\
From the available visible final states, a number of observables are constructed that are then input into the TMVA package~\cite{TMVA2007}, which handles a Multi-Variate Analysis (MVA). Among the different analysis techniques, we employed the Boosted Decision Tree (BDT) which is largely used by the LHC experimental collaborations. 
\begin{figure}
\includegraphics[width=0.356\textwidth]{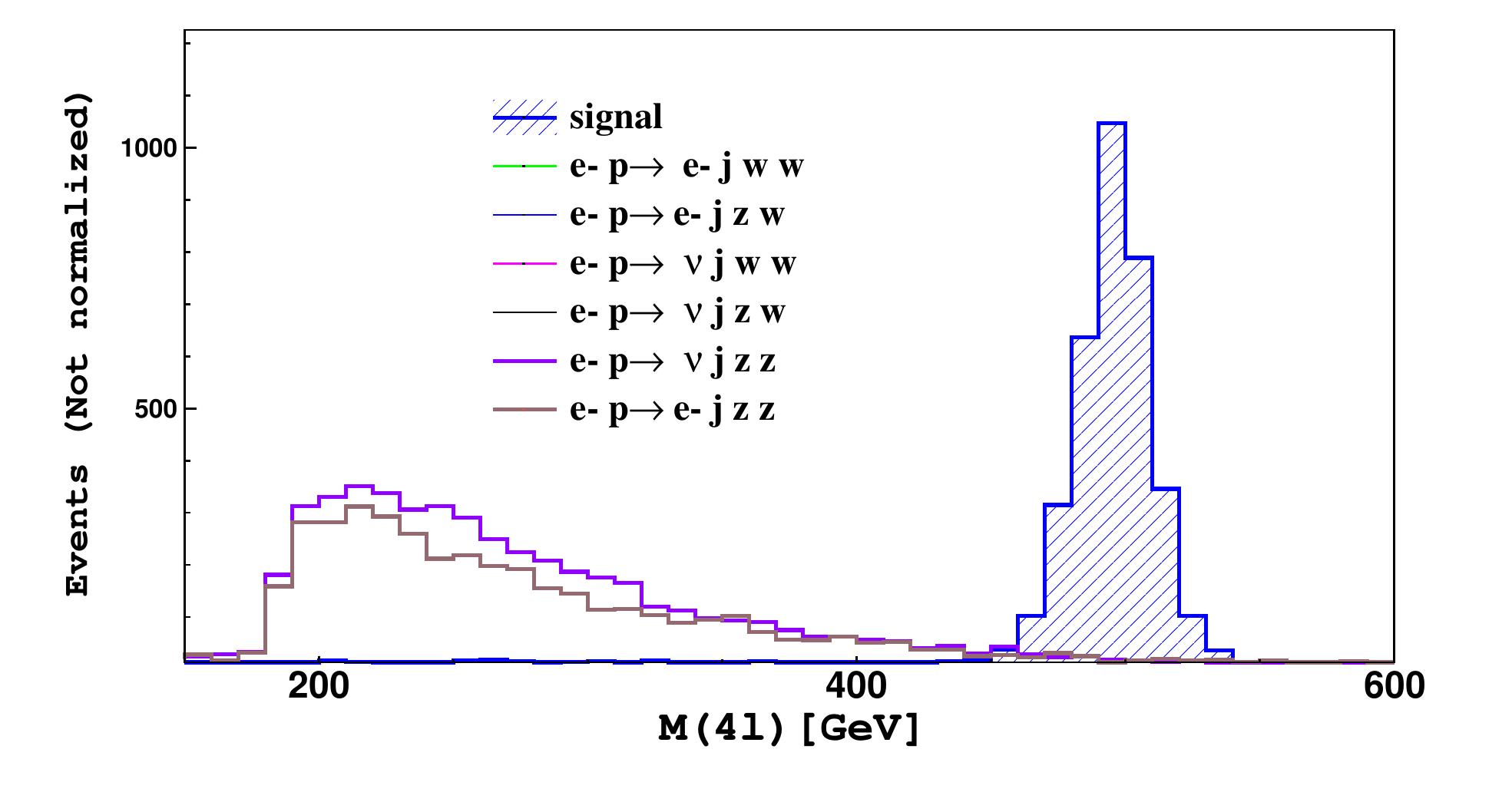}
\includegraphics[width=0.335\textwidth]{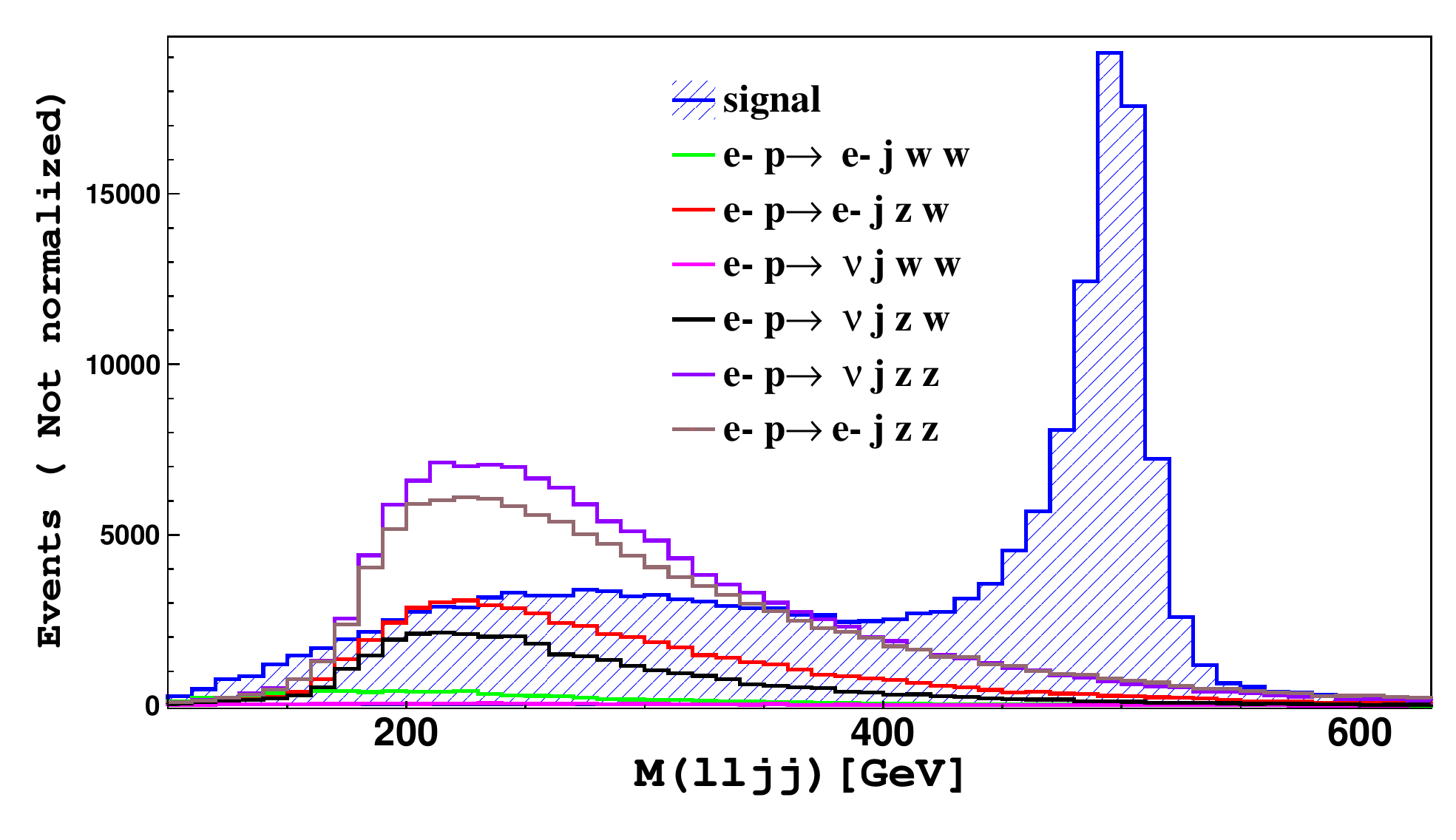}
\includegraphics[width=0.30\textwidth]{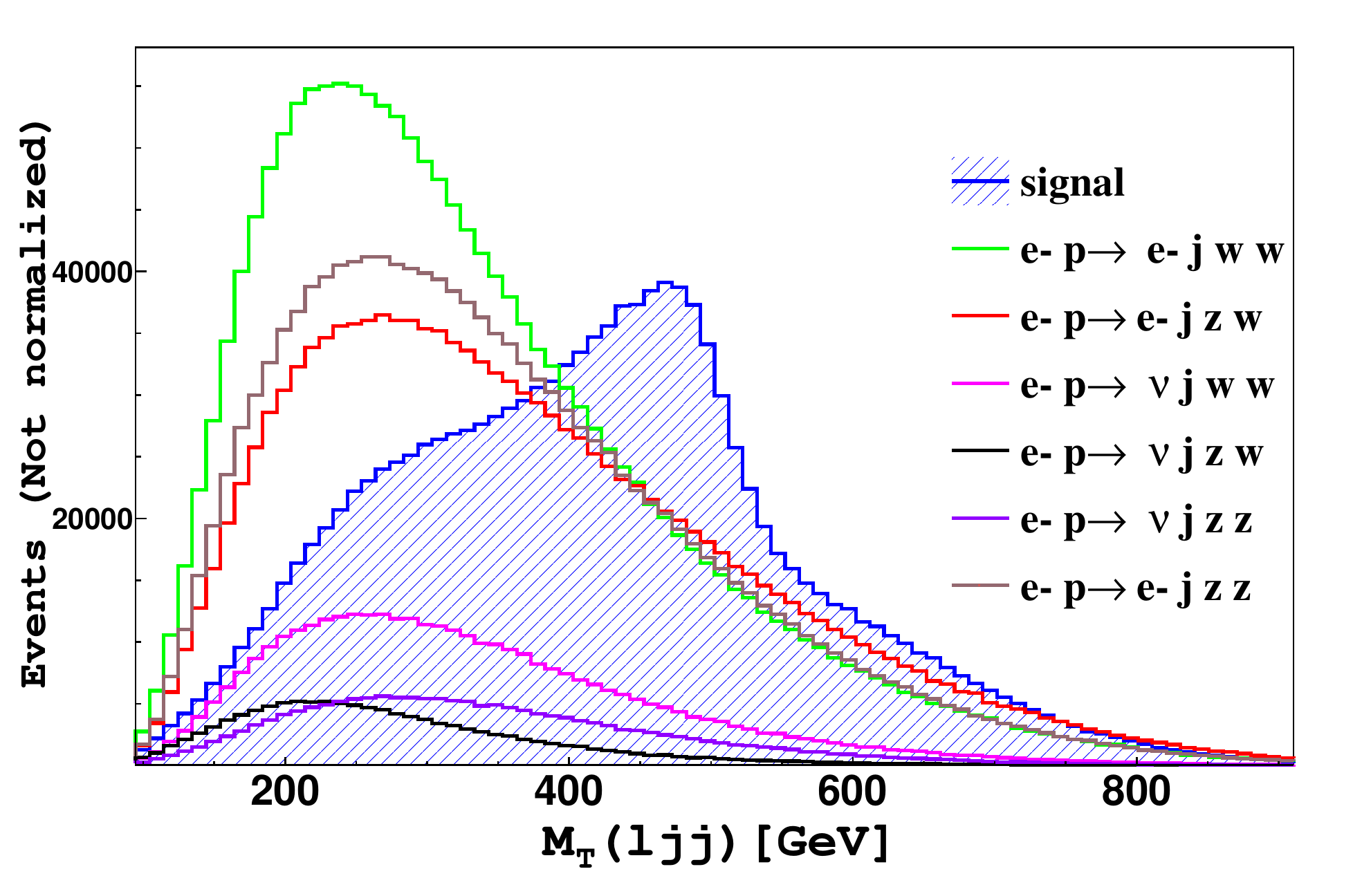}
\caption{The most relevant observables as ranked by the BDT analysis for the three signal channels $\mu_{\ell\ell}^Z$ (left),  $\mu_{\ell q}^Z$ (middle) and  $\mu_{\ell q}^{W}$ (right) with a signal benchmark point defined by $m_{h_2} = 500$ GeV and $\sin \alpha =0.2$. The variable in the left plot is the invariant mass of four final state leptons. The variable in the middle plot is the invariant mass of two final state jets and two final state leptons. The variable in the right plot is the transverse mass of the lepton and the two jets in the final state.}
\label{fig:var}

\includegraphics[width=0.356\textwidth]{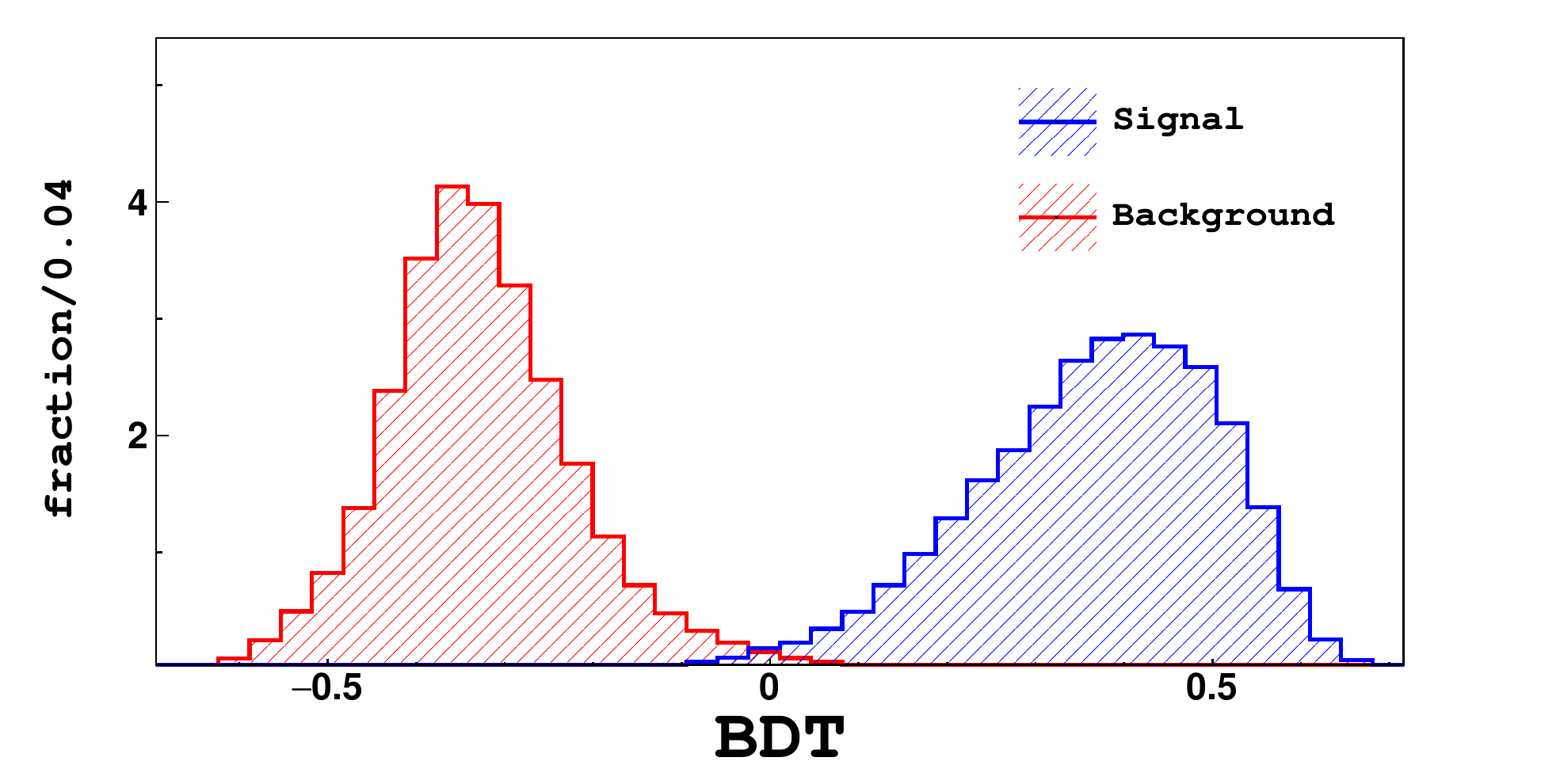}
\includegraphics[width=0.335\textwidth]{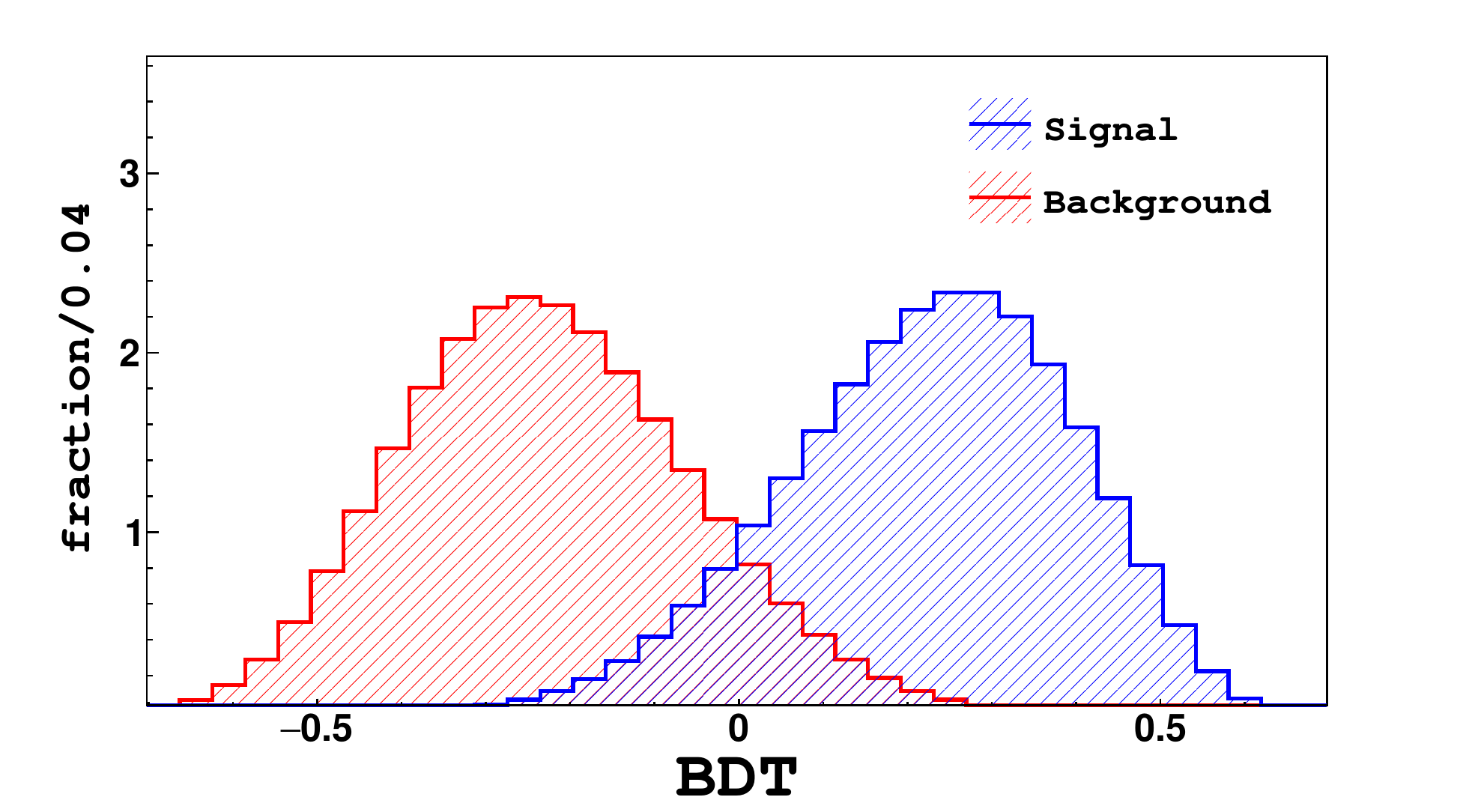}
\includegraphics[width=0.30\textwidth]{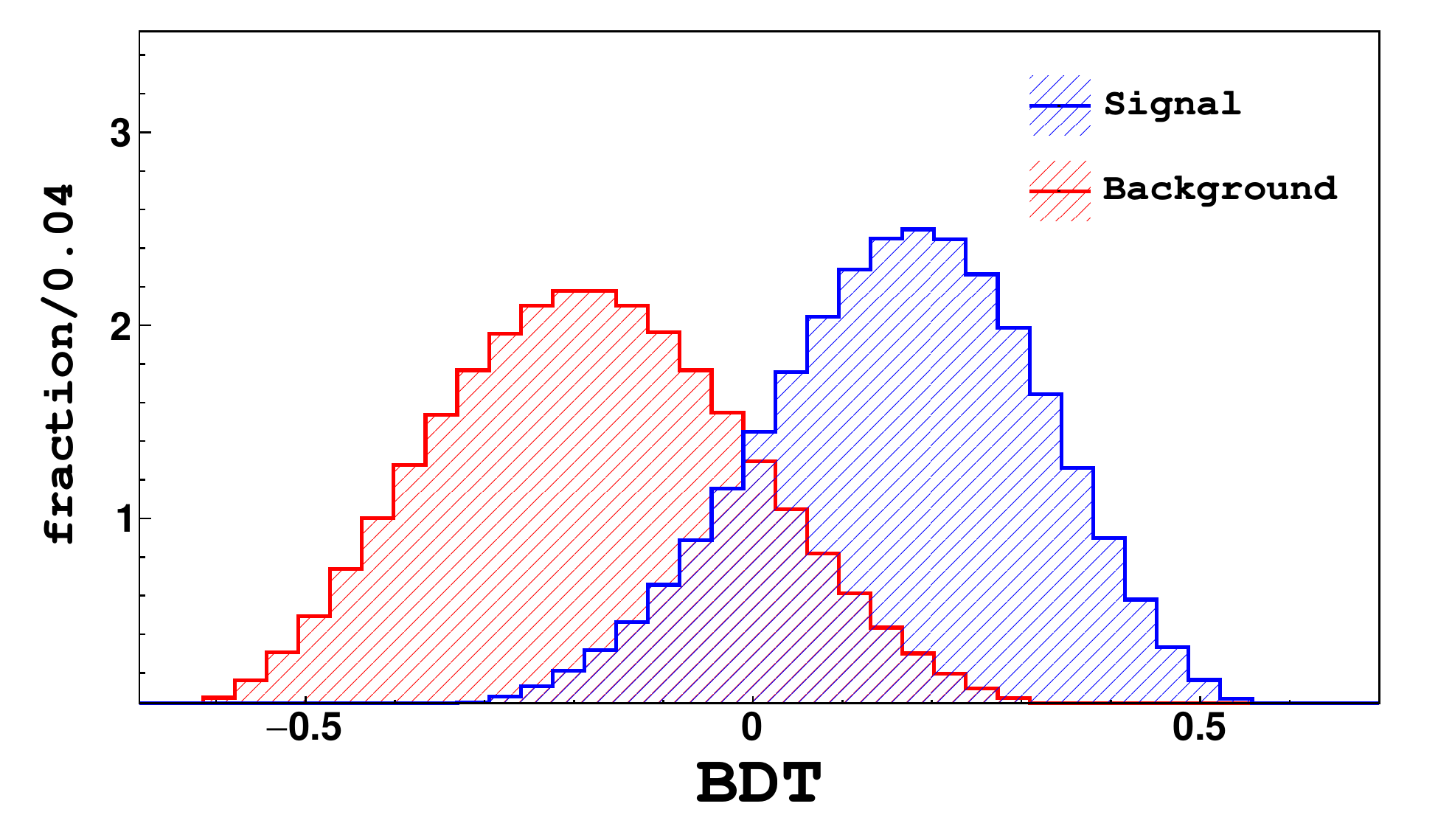}
\caption{The BDT distributions for the three signal channels $\mu_{\ell\ell}^Z$ (left),  $\mu_{\ell q}^Z$ (middle), and  $\mu_{\ell q}^{W}$ (right) with $m_{h_2} = 500$ GeV and $\sin \alpha =0.2$. }
\label{fig:BDT}

\includegraphics[width=0.356\textwidth]{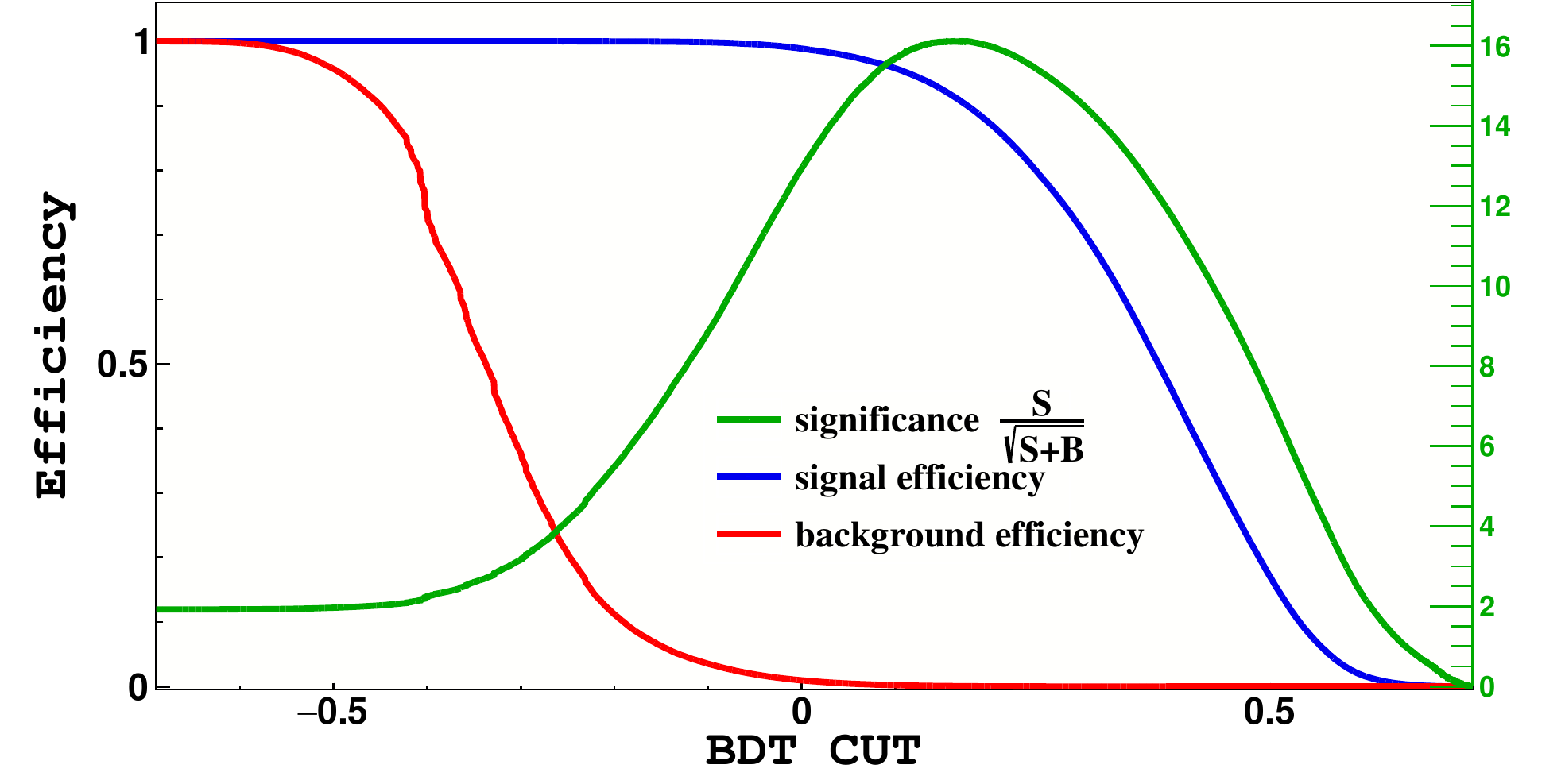}
\includegraphics[width=0.335\textwidth]{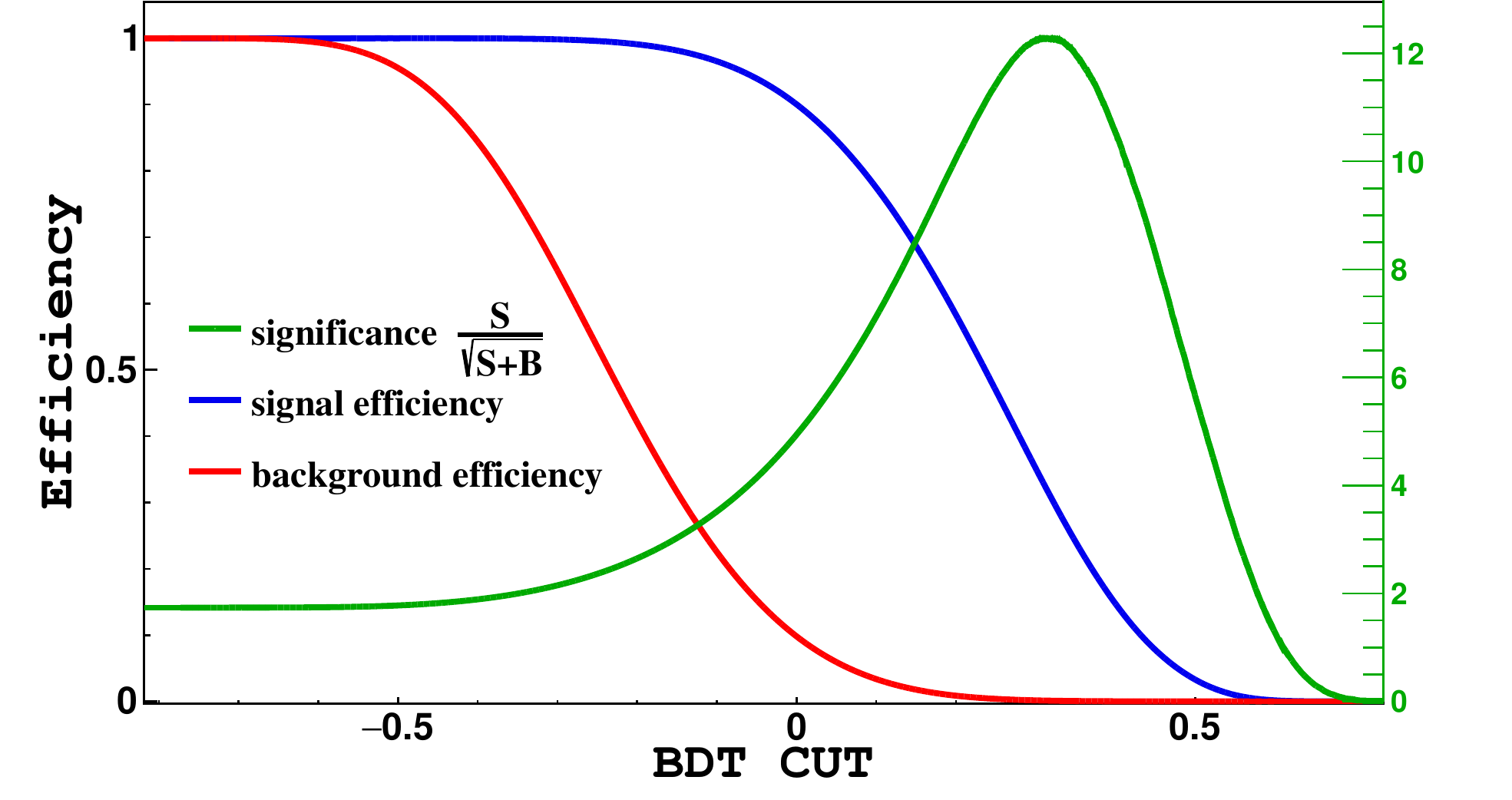}
\includegraphics[width=0.30\textwidth]{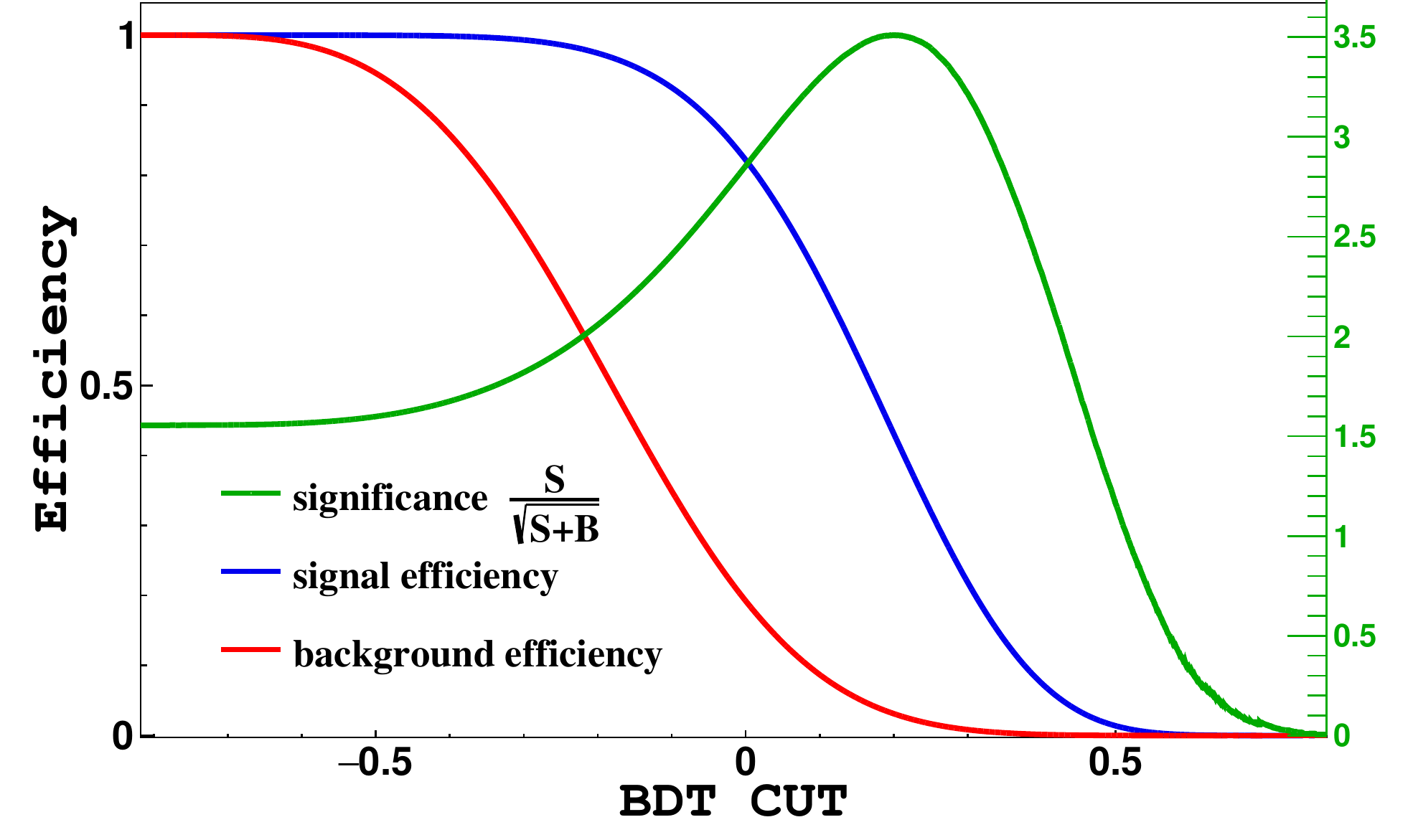}
\caption{Cut efficiency  and the relevant significance distributions for the three signal channels $\mu_{\ell\ell}^Z$ (left),  $\mu_{\ell q}^Z$ (middle), and  $\mu_{\ell q}^{W}$ (right) with $m_{h_2} = 500$ GeV and $\sin \alpha = 0.2$.}
\label{fig:eff}
\end{figure}

\paragraph{The fully leptonic final state, $\mu_{\ell\ell}^Z$:}
This signal channel consists of two lepton pairs $\ell_\alpha^+ \ell_\alpha^- \ell_\beta^+ \ell_\beta^-$ with the lepton flavours $\alpha,\beta\in \{e,\,\mu\}$.
Conservatively, we considered here only electrons and muons but, in principle, reconstructing tau leptons is also possible and may enhance the significance of this channel.
We observed that the final state events are characterized by a highly boosted beam-remnant jet with large positive $\eta$ values. 

The mass of $h_2$ in the range investigated here is always larger than $2m_Z$, thus we can require the two $Z$ bosons to be produced on shell and to decay leptonically. 
From the possible combinations of same flavour and opposite sign leptons, one should recover the invariant masses of both pairs compatibly with $m_Z$.
Nevertheless, for our pre-selection before the actual MVA, we do not explicitly require the lepton pairs to reconstruct the $Z$ boson peaks within a given mass window but simply collect, among all possible leptons in the events, the electron or muon pairs that are closest to the $Z$ rest mass. Final state leptons are thus grouped into three categories: $4\mu$, $2e 2\mu$ or $4e$. 
The main source of irreducible backgrounds in this case is given by $\nu jZZ$ and $e^- j ZZ$, while the other backgrounds that contain $W$ bosons are suppressed during the MVA process by exploiting the different positions of the peaks in the invariant mass distributions of the lepton pairs. \\
We take advantage of the full power of BDT algorithm in distinguishing between signal and background events, feeding it with 42 kinematical observables. 
The most relevant observables according to the BDT ranking are, as expected, the invariant mass of $h_2$ from $4\mu$, $2e2\mu$ and $4e$ respectively, as well the reconstructed invariant mass of $Z$ boson. 
As an example, the four-lepton invariant mass distribution for the signal and the background samples is shown in fig.\ \ref{fig:var} for the benchmark point defined by $m_{h_2} = 500 $ GeV and $\sin \alpha = 0.2$. 
We checked that changing the pre-selection cut of the beam remnant jet $P_T$ from 10 GeV to 20 GeV affects the final result for the leptonic channel, which is statistically dominant, only by about 3\%.

\paragraph{The semileptonic final state, $\mu_{\ell q}^{Z}$:}
In this channel the two leptons and two jets can be paired up to the $Z$ boson mass. The invariant mass of the two Z boson candidates in turn reconstructs to $m_{h_2}$.
The analysis strategy follows quite closely the one of the fully leptonic final state described above.
In particular, we collect among all possible leptons and jets in the events, the lepton and jet pairs that are closest to the $Z$ mass and we further organize the events into two categories according to the flavour of the lepton pair.
The most relevant irreducible background for this channel stems from $\nu jZZ$ and $e^- j ZZ$ but further sizable contributions exist from processes with at least one $W$ boson. 
The reconstructed invariant mass of the $h_2$, the pseudo rapidity distribution of leptons and the angular separation between the leptons and the reconstructed $Z$, $\Delta R(Z,\ell)$, are classified by the BDT algorithm as the most relevant observables in distinguishing signal from backgrounds. The invariant mass distribution of the lepton-jet system for the signal and the background samples is shown in fig.\ \ref{fig:var} for the benchmark point defined by $m_{h_2} = 500$ GeV and $\sin \alpha = 0.2$.
 
Differently from the $\mu_{\ell \ell}^{Z}$ above, the larger cross section allows to access heavier masses with respect to the fully leptonic final state and, as such, makes this process suitable for searches of heavy scalars with larger masses.

\paragraph{The semileptonic final state, $\mu_{\ell q}^{W}$:}
This signal channel is much more difficult to reconstruct compared to the first two due to the final state neutrino which escapes from the detector and makes it impossible to fully reconstruct the $h_2$ system. \\
For our pre-selection we select in each event, among all jets with highest momentum, the two jets with the reconstructed invariant mass that is closest to the $W$ boson mass, and, among all possible leptons, that with the highest momentum that together with the missing energy reconstructs more closely the transverse mass of the second $W$. The main discriminating variable here is the transverse mass of $h_2$ which is peaked around the rest mass of $h_2$ and has a flat tail, due to the missing energy contribution. This distribution is shown if fig.\ \ref{fig:var} for a particular benchmark point. 
Further relevant observables are the invariant mass of the $(W+l)$ system, the $\eta (W,l)$ and $\eta (\ell)$ distributions. \\
Here the usage of MVA, and especially the BDT, is found to be extremely useful, with respect to standard cut-based analysis, in exploiting the differences between signal and background distributions.

\begin{table}
\centering
\begin{tabular}{clc}
\hline
Nr. & final state & $\sigma_{\rm LHeC}$ [fb] \\
\hline\hline
1 & $ e^- j WW$ & 23.0 \\
2 & $e^- j ZW^+$ & 4.16 \\
3 & $e^- j ZZ$ & 0.1  \\
4 & $\nu j WW$ & 10.4  \\
5 & $\nu j ZW^-$ & 8.0 \\
6 & $\nu j ZZ$ & 2.4 \\
\hline
\end{tabular}
\caption{The SM background processes considered in this analysis. The samples have been produced with the following cuts: $P_T(j)> 10$ GeV,  $P_T(l)> 2$ GeV and $|\eta(j/l)| < 4.5$.}
\label{tab:bkg}
\end{table}

\subsection{Results}
\begin{figure}
\centering
\includegraphics[width=10cm,height=5.5cm]{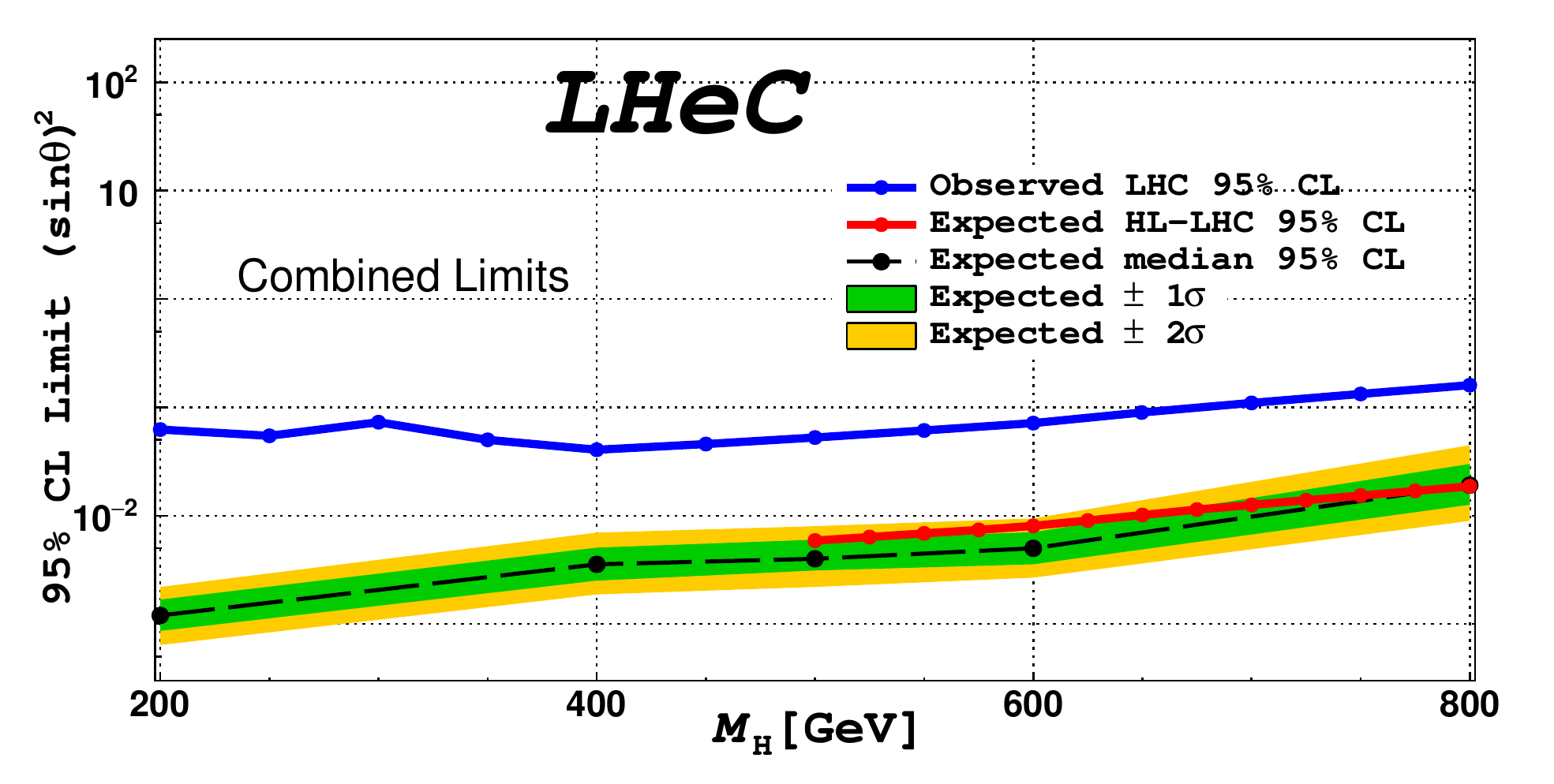}
\caption{Combined limit for the three signal channels, including a systematic uncertainty of $2\%$. The blue line represents the current LHC limit at $95\%$ CL as extracted from \cite{Sirunyan:2018qlb}, the red line the forecast of the HL-LHC sensitivity via $h_2 \to ZZ$ searches from ref.\ \cite{CMS:2019qzn}.}
\label{fig:limit_combined}
\end{figure}

We employ the BDT method to perform the multivariate analysis. The discriminating power of the BDT relies on the fact that the signal and the background may be characterised by different features that can be entangled together into several distributions.
When these features are not clearly manifest in some specific observables, it could be difficult, in principle, to identify the most relevant distributions able to efficiently separate the signal events from the background ones.
This machine learning technique is based on a set of decision trees where each tree yields a binary output depending on the fact that an event is classified as signal-like or background-like during the training session.
The main advantage of the algorithm consists on the possibility to combine together several discriminating variables (in our study we have employed 42 kinematical observables for both signal and background events) into a single and more effective discriminator, the BDT variable, and thus to reach a higher significance with respect to standard methods.
  
Figure \ref{fig:BDT} shows the BDT distributions for the three channels $\mu_{\ell\ell}^Z$ (left), $\mu_{\ell q}^Z$ (middle), and $\mu_{\ell q}^{W}$ (right) for the benchmark point given by $m_{h_2} = 500$ GeV and $\sin \alpha = 0.2$. The BDT discriminator ranges from $-1$ to $1$: the events with discriminant value near $1$ are classified as signal-like events (blue distribution) and those near $-1$ are considered as background-like events (red distribution).  
The optimization of the signal/background cut, as a function of the BDT variable, has been performed using the TMVA and expressed in terms of the significance $\frac{S}{\sqrt{S+B}}$. Figure \ref{fig:eff} shows the cut efficiency for the three channels $\mu_{\ell\ell}^Z$ (left),  $\mu_{\ell q}^Z$ (middle), and  $\mu_{\ell q}^{W}$ (right) for $m_{h_2} = 500$ GeV and $\sin \alpha = 0.2$. 
For $\mu_{\ell\ell}^Z$ channel, by requiring BDT $>0.163$, we can reach significance $\frac{S}{\sqrt{S+B}}= 16.1\sigma$ with signal efficiency 0.91 and background rejection efficiency of $6.4 \times 10^{-4}$, for the channel $\mu_{\ell q}^Z$ with BDT$>0.313$ we obtain a significance of $12.28\sigma$ with signal efficiency 0.4 and background efficiency $1.2 \times 10^{-3}$. Finally, for the channel $\mu_{\ell q}^{W}$, with BDT$>0.23$ one can get a significance of $3.5\sigma$ with signal efficiency 0.43 and background efficiency 0.034.

The combined sensitivity is derived from the BDT distributions of the above described analyses and for each benchmark mass. As stated above, we included a the systematic uncertainty on the background of $2\%$ and we used the Higgs Analysis-Combined Limit tool~\cite{combine}. To extract the limit we preformed a frequentist test which uses the profile likelihood as test statistics. In addition to the parameters of interest, such as the total cross section and the integrated luminosity, we include a nuisance parameter for background only of $2\%$ as a logarithmic-normal distribution to account for the unknown systematic uncertainty of the future LHeC. In fig.\ \ref{fig:limit_combined} we show the $95\%$ CL expected median limit on the squared sine of the mixing angle, as well the error bands for 1 and 2 sigma. 
Due to the different efficiencies, branching fractions and the relevant backgrounds, each final state contributes differently depending on the mass of the heavy scalar. As an example, we find that the $\mu_{\ell\ell}^Z$ channel is the most sensitive one in the mass range $200-500$ GeV, while the $\mu_{\ell q}^Z$ channel is sensitive in the higher mass regime. The current LHC limit (red dashed line) at $95\%$ CL has been extracted from \cite{Sirunyan:2018qlb}, where the search has been performed for heavy scalars over the mass range of 130 GeV to 3 TeV at a centre of mass energy of $13$ TeV and 35.9/fb of integrated luminosity. 
In particular, the $Z$ boson pair decay channel has been investigated in the final state objects $4l, 2l2j$ and $2l2\nu$.
It is clear that the sensitivity of the LHeC is better than the current LHC one by about two orders of magnitude in the low mass regime up to one order of magnitude in the high mass region.
As an example, the expected 2$\sigma$ median sensitivity of the LHeC to $\sin^2 \alpha$ for the mass $m_{h2}=500$ GeV can be as small as $4\times 10^{-3}$. 

For masses up to about 1 TeV this sensitivity is comparable to the forecast of the HL-LHC sensitivity via $h_2 \to ZZ$, which we extract  from ref.\ \cite{CMS:2019qzn}.
It is worthy of note that the LHeC's higher sensitivity for heavy scalars with masses of only a few hundred GeV is complementary to the higher sensitivity of the HL-LHC at masses on the TeV scale.

\section{Conclusion}
Precision measurements of the Higgs boson properties are very important due to its possible role as portal to BSM sectors. Present searches at the LHC are compatible with additional heavy scalar particles that mix on the percent level with the SM Higgs boson.
We have shown that the prospects of discovering such heavy scalars at the LHeC are very promising and complementary to the searches at the LHC, where the notorious SM backgrounds and systematic uncertainties make discovery difficult.
Using multivariate techniques and by exploiting three of the most promising decay channels of a heavy Higgs, we find that the LHeC can access heavy scalar bosons with masses between 200 and 800 GeV and scalar mixings as small as $\sin^2 \alpha \sim 10^{-3}$. \\
We also pointed out that many other interesting channels exist that may allow to test the properties and the origin of the heavy Higgs boson. Among these, searches for (semi) invisible decays, di-higgs and the di-top final states may successfully exploit the cleaner environment offered by the promising future LHeC.

The superb reach to small scalar mixings for masses below one TeV makes the LHeC complementary to the possible reach of the HL-LHC for larger masses. We therefore conclude, that this machine is uniquely suited in order to discover and study possible scalar bosons with masses ${\cal O}(100)$ GeV.

\paragraph{Acknowledgements:} 
The authors acknowledge support from the LHeC Study Group.
A. Hammad would like to thank Waleed Esmail and Ashraf Kasem for the fruitful discussions.
O. Fischer received funding from the European Unions Horizon 2020 research and innovation program under the Marie Sklodowska-Curie grant agreement No 674896 (Elusives). 
A.Hammad is supported by the Swiss National Science Foundation.

\appendix

\section{Multi-variate analysis with a Boosted Decision Tree}
For our analysis we employed the TMVA package~\cite{TMVA2007} which incorporates a Boosted Decision Tree (BDT). 
This algorithm allowed us to exploit several kinematical distributions, listed in tab.\ \ref{table:cut}, for each of the three considered signal channels.
The signal and background samples are divided into a training and an analysis set.
The BDT is first trained on the training sample in order to construct the classifier that assesses whether an event is from a signal or from a background process.
The analysis sample is then used to test the final classifier, which is obtained in the following way:
\begin{enumerate}
\item A root node is created from an initial number of sample events.
\item The algorithm finds the optimal threshold for a given kinematical observable that gives the best separation between signal and background
Afterwards it separates the node into two branches, one containing mostly signal and the other mostly background events.
\item For events that are neither signal- nor background-like, the BDT continues iterating through the input observables to decide on the nature of the unclear events. 
\item Considering each branch as a new node, the algorithm goes through steps 2 and 3 and keeps repeating it until a given number of final branches (called leaves) are obtained, which correspond to true signal or pure background.
\end{enumerate}
A given brach will define a next node according to the purity coefficient
\begin{equation}
P = \frac{signal\ events}{(signal\ events+ backgrounds\ events)}
\end{equation}
and the splitting criterion of a given branch, obtained by maximising the signal/background separation, is defined by the Gini index
\begin{equation}
G_i = \sum^n_i W_i P( 1-P )\,,
\end{equation}
where $W_i$ is the weight of the events $i$. 
The observable with the highest discrimination power (of signal from background events) is obtained by maximising the difference between the Gini index of the parent node and the sum of Gini indices of the two daughter nodes.

The training sample is reweighted such that the decision tree starts with a maximal Gini index, where signal events are equal to background events (i.e.\ $P = 0.5)$. A leaf with purity greater than $0.5$ is called a signal leaf, otherwise it is a background leaf. Several trees can combined together into a so-called forest and the final BDT output discriminator is chosen such that statistical fluctuations are minimized. The BDT discriminator ranges between $-1$ and $1$ corresponding to pure background and pure signal, respectively.

\section{Statistical evaluation of the BDT output}
For the statistical treatment of our BDT analysis we used the Higgs Analysis-Combined Limit tool~\cite{combine}, which allows for different statistical procedures.
From the available options we chose to use a frequentist test of profile Likelihood ratios as test statistics.
Besides the total cross section and the integrated luminosity, we included among the parameters of the test statistics a
nuisance parameter for the background with a relative strength of $2\%$ to account for the systematic uncertainties at LHeC as quoted by the Conceptual Design Report \cite{AbelleiraFernandez:2012cc}.
The tool then computes the probability of finding the observed (simulated) data incompatible with the prediction for a given hypothesis, the $p$-value.
The expected value of finding the number of events in the $i$th bin of the BDT distribution is given by
\begin{equation}
E[n_i] = \mu S_i + B_i\,,
\end{equation}
where the parameter $\mu$ is called the signal strength. 
The signal is excluded at $(1-\alpha)$ confidence level if
\begin{equation}
CL_s = \frac{P\left( q(\mu) | \mu S + B \right)}{P\left( q(\mu) | B \right)} < \alpha ,
\end{equation}
where $q(\mu)$ is the profile log likelihood.
Finally, the error bands can be obtained by
\begin{equation}
{\rm Band}_{(1-\alpha)}= \hat{\mu} \pm \frac{\sigma\Phi^{-1}\left( 1-\alpha \right)}{N}\,,
\end{equation}
where $\hat \mu$ is the estimated expected median and $\Phi^{-1}$ is the cumulative distribution.
If we restrict the number of events for the signal and the background to be large and ignore the correlation effect between bins, the significance can be described by the following formula
\begin{equation}
\sigma_{\rm stat+syst} =
\Bigg[ 2 \bigg( (N_s + N_b) {\rm ln} \frac{(N_s+N_b)(N_b+\sigma_b^2)}{N_b^2+(N_s+N_b)\sigma_b^2} - \frac{N_b^2}{\sigma_b^2} {\rm ln}(1+ \frac{\sigma_b^2 N_s}{N_b(N_b+\sigma_b^2) } )\, \bigg)\, \Bigg]^{1/2}
\label{eqn:sgf2}
\end{equation}
with $N_s$, $N_b$ being the number of signal and background events, respectively, and $\sigma_b$ parametrising the systematic uncertainty.

\section{Variables ranking}
In the tables below the following definitions have been adopted:
$M$ is the invariant mass, $P_T$ is the transverse momentum, $\eta$ is the pesudorapidity, $\Delta R$ is the angular separation between two isolated objects and is defined as $\Delta R = \sqrt{\Delta\eta^2 +\Delta\Phi^2}$, $M_T$ is the transverse mass given by$M^2_T = \left(\sqrt{M^2(f)+P^2_T(f)}+|P^{miss}_T(f)| \right)^2 -\left(\vec{P_T}(f)+\vec{P}_T^{miss} \right)^2 $.

\begin{table}[h]

\begin{minipage}{.33\linewidth}
%      \caption*{$h_2\to ZZ\to 4l$}
 %  \centering
  \resizebox{\textwidth}{!}{
\begin{tabular}{|c|c|c|}
\hline
\multicolumn{3}{|c|}{$h_2\to ZZ\to 4l$}\\
\hline
Ranking  & Observable & Importance \\ [0.5ex]
\hline\hline
1 & $M(4\mu)$ & $6.400$   \\
 \hline
 2 & $M(2e2\mu)$ & $5.989$   \\
 \hline
 3 & $M(4e)$ & $5.715$   \\
 \hline
  4 & $P_T(2\mu)$ & $5.028$   \\
 \hline
  5 & $\Delta R(e,\mu) $ & $4.887$   \\
 \hline
  6 & $\Delta R(\mu,\mu)$ & $4.342$   \\
 \hline
   7 & $\Delta R(2\mu,2\mu) $ & $3.797$   \\
 \hline
    8 & $P_T(2e)_{Z_1} $ & $3.703$   \\
 \hline
    9 & $\Delta R(\mu,\mu) $ & $3.659$   \\
 \hline
    10 & $\Delta R(e,e) $ & $3.659$   \\
 \hline
    11 & $\eta(4e) $ & $3.600$   \\
 \hline
    12 & $P_T(j)_{beam}$ & $3.382$   \\
 \hline
    13 & $M(\mu\mu)$ & $3.374$   \\
 \hline
   14 & $\eta(2\mu)$ & $3.274$   \\
 \hline
   15 & $\Delta R(e,\mu)$ & $3.031$   \\
 \hline
   16 & $\eta(2e)$ & $3.001$   \\
 \hline
   17 & $P_T(e,\mu)$ & $2.940$   \\
 \hline
   18 & $\Delta R(e,e)$ in case of $4e$ & $2.562$  \\ 
 \hline
   19 & $P_T(4\mu)$ & $2.487$   \\
 \hline
   20 & $\Delta R(4e,j)$ & $2.404$   \\
 \hline
   21 & $\eta(2e,2\mu)$ & $2.404$   \\
 \hline
   23 & $\Delta R (2e2\mu, j)$ & $2.207$ \\  
 \hline
   24 & $\eta(4\mu)$ & $2.199$ \\  
 \hline
   25 & $\Delta R(4\mu,j)$ & $1.928$ \\  
 \hline
   26 & $\eta(2e)$ & $1.775$ \\  
 \hline
   27 & $\eta(e,m)$ & $1.711$ \\  
 \hline
   28 &MET& $1.652$ \\  
 \hline
    29 &$\eta(2\mu)$& $1.628$ \\  
 \hline
    30 &$P_T(4e)$& $1.619$ \\  
 \hline
    31 &$P_T(2e2\mu)$& $1.582$ \\  
 \hline
    32 &$\Delta R (2e,2e)$& $1.012$ \\  
 \hline
    33 &$\Delta R (2e,2\mu)$& $.07312$ \\  
 \hline
\end{tabular}}
\end{minipage}%
\begin{minipage}{.33\linewidth}
     %\centering
%        \caption*{$h_2 \to ZZ \to 2l2j$}
        \resizebox{\textwidth}{!}{
\begin{tabular}{|c|c|c|}
\hline
\multicolumn{3}{|c|}{$h_2\to ZZ\to 2l2j$}\\
\hline
Ranking  & Variable &Importance \\ [0.5ex]
\hline\hline
1 & $M(2\mu 2j)$ & $5.534$   \\
 \hline
 2 & $M(2e2j)$ & $4.516$   \\
 \hline
 3 & $\eta(e^-)$ & $4.372$   \\
 \hline 
 4 & $P_T(\mu^- \mu^+)$ & $4.305$   \\
 \hline  
 5 & $P_T(jj)_z$ & $3.942$   \\
 \hline  
 6 & $\eta(\mu^-\mu^+jj)$ & $3.885$   \\
 \hline  
 7 & $\Delta R(e^-e^+)_z$ & $3.850$   \\
 \hline  
 8 & $P_T(e^-e^+)_z$ & $3.793$   \\
 \hline  
 9 & $\Delta R(\mu^-\mu^+)_z$ & $3.383$   \\
 \hline  
 10 & $\eta(e^-e^+jj)$ & $3.302$   \\
 \hline  
 11 & $MET$ & $3.261$   \\
 \hline  
 12 & $P_T(j)_{beam}$ & $2.938$   \\
 \hline  
 13 & $\eta(j)_{beam}$ & $2.763$   \\
 \hline  
 14 & $M(\mu^-\mu^+)_z$ & $2.737$   \\
 \hline  
 15 & $P_T(\mu^+)$ & $2.707$   \\
 \hline  
 16 & $\eta(j_2)_z$ & $2.696$   \\
 \hline  
 17 & $\eta(jj)_z$ & $2.585$   \\
 \hline  
 18 & $P_T(\mu^-\mu^+jj)$ & $2.520$   \\
 \hline  
 19 & $\Delta R(e^-e^+jj,j_{beam})$ & $2.506$   \\
 \hline  
 20 & $\Delta R(\mu^-\mu^+jj,j_{beam})$ & $2.397$   \\
 \hline  
 21 & $P_T(\mu^-\mu^+)_z$ & $2.394$   \\
 \hline  
 22 & $P_T(e^+)$ & $2.306$   \\
 \hline  
 23 & $M(jj)_z$ & $2.257$   \\
 \hline  
 24 & $PT(e^-e^+jj)$ & $2.218$   \\
 \hline  
 25 & $\eta(j_1)_z$ & $2.171$   \\
 \hline  
 26 & $\eta(\mu^-\mu^+)_z$ & $2.161$   \\
 \hline  
 27 & $M(e^-e^+)_z$ & $1.931$   \\
 \hline  
 28 & $P_T(e^-)_z$ & $1.930$   \\
 \hline  
 29 & $P_T(j_2)_z$ & $1.920$   \\
 \hline  
 30 & $\eta(e^+)_z$ & $1.701$   \\
 \hline  
 31 & $P_T(j_1)_z$ & $1.694$   \\
 \hline  
 32 & $\eta(e^-e^+)_z$ & $1.639$   \\
 \hline  
 33 & $\eta(\mu^-\mu^+)_z$ & $1.638$   \\
 \hline  
 34 & $\eta(\mu^+)_z$ & $1.633$   \\
 \hline  
 35 & $\Delta R(j,j)_z$ & $1.487$   \\
 \hline  
 36 & $\Delta R(jj,\mu^-\mu^+)$ & $1.463$   \\
 \hline  
 37 & $\Delta R(jj,e-e^+)$ & $1.462$   \\
 \hline  
\end{tabular}}
\end{minipage}%
\begin{minipage}{.33\linewidth}
     %\centering
%        \caption*{$h_2 \to W^\pm W^\mp \to 2jl\nu_l$}
        \resizebox{\textwidth}{!}{
\begin{tabular}{|c|c|c|}
\hline
\multicolumn{3}{|c|}{$h_2\to W^\pm W^\mp \to 2jl\nu_l$}\\
\hline
Ranking  & Variable &Importance \\ [0.5ex]
\hline\hline
1 & $M_T(\mu j_wj_w)$ & $18.24$   \\
 \hline  
 2 & $P_T(jj)_w$ & $17.96$   \\
 \hline 
 3 & $P_T(j_1)_w$ & $14.66$   \\
 \hline 
 4 & $M(j_wj_wj_{beam})$ & $13.30$   \\
 \hline 
 5 & $\eta(e)$ & $12.74$   \\
 \hline 
 6 & $M(\mu j_wj_w)$ & $11.77$   \\
 \hline 
 7 & $P_T(\mu)$ & $9.395$   \\
 \hline 
 8 & $M(\mu j_wj_wj_{beam})$ & $9.285$   \\
 \hline 
 9 & $M_T(e j_wj_w)$ & $8.865$   \\
 \hline 
 10 & $\eta(e j_wj_w)$ & $7.830$   \\
 \hline 
 11 & $P_T(j_{beam})$ & $7.590$   \\
 \hline 
 12 & $P_T(j_2)_w$ & $5.969$   \\
 \hline 
 13 & $\eta(j_2)_w$ & $5.740$   \\
 \hline 
 14 & $\eta(\mu j_wj_w)$ & $5.275$   \\
 \hline 
 15 & $MET$ & $4.724$   \\
 \hline 
 16 & $P_T(ej_wj_wj_{beam})$ & $4.329$   \\
 \hline 
 17 & $\eta(j_wj_wj_{beam})$ & $4.277$   \\
 \hline 
 18 & $P_T(e)$ & $4.176$   \\
 \hline 
 19 & $M(ej_wj_w)$ & $4.137$   \\
 \hline 
 20 & $\eta(\mu j_wj_wj_{beam})$ & $3.658$   \\
 \hline 
 21 & $M(e j_wj_wj_{beam})$ & $3.544$   \\
 \hline 
 22 & $\eta(j_{beam})$ & $3.357$   \\
 \hline 
 23 & $\eta(\mu)$ & $2.970$   \\
 \hline 
 24 & $\eta(j_1)_w$ & $2.570$   \\
 \hline 
 25 & $M(jj)_w$ & $2.527$   \\
 \hline 
 26 & $\eta(ej_wj_wj_{beam})_w$ & $2.346$   \\
 \hline 
 27 & $P_T(j_wj_wj_{beam})_w$ & $2.019$   \\
 \hline 
 28 & $P_T(\mu j_wj_w)_w$ & $2.007$   \\
 \hline 
 29 & $P_T(ej_wj_w)_w$ & $1.712$   \\
 \hline 
 30 & $P_T(\mu j_wj_w j_{beam})_w$ & $1.552$   \\
 \hline 
 31 & $\eta(jj)_w$ & $0.6948$   \\
 \hline 
\end{tabular}}
\end{minipage}

\caption{Variables ranking for $m_{h_2} = 500$ GeV, the importance is in percent. For the four lepton final states also considered were the 9 observables $M(Z)$, and $P_T(Z)$ for the two different $Z$ boson candidates, and $\eta(j)$, which resulted in a BDT ranking of 0.}
\label{table:cut}

\end{table}

%%%%%%%%%%%%%%%%%%%%%%%%%%%%%%%%%%%
\bibliographystyle{unsrt}

\end{document}